\documentclass[aps,nofootinbib,superscriptaddress,longbibliography]{revtex4-2}
\usepackage{amsmath,amssymb,graphicx,color}
\usepackage[section]{placeins}
\usepackage{algorithm}
\usepackage{algpseudocode}
\usepackage{tikz}
\usepackage{tikz-network}
\graphicspath{{./}}

\newcommand{\nbverts}{N}
\newcommand{\nbstates}{M}
\newcommand{\rSI}{\beta}
\newcommand{\rIS}{\gamma}
\newcommand{\occup}{n}
\newcommand{\ladderup}{\sigma^+}
\newcommand{\ladderdown}{\sigma^-}
\newcommand{\HSIS}{H_\mathrm{SIS}}
\newcommand{\hSIS}{\mathcal{H}_\mathrm{SIS}}
\newcommand{\htpar}{\alpha} 

\begin{document}
\title{Restoring detailed balance in non-Hermitian Markov processes}
\author{Tim Van Wesemael}
\affiliation{BionamiX, Department of Data Analysis and Mathematical Modelling, Ghent University, 9000 Ghent, Belgium}
\author{Gilberto Nakamura}
\affiliation{FACOM, Universidade Federal de Uberlândia, Uberlândia 38400-902, Brazil}
\affiliation{Instituto de Física de São Carlos, Universidade de São Paulo, São Carlos 13566-590, Brazil}
\author{Jan Baetens}
\affiliation{BionamiX, Department of Data Analysis and Mathematical Modelling, Ghent University, 9000 Ghent, Belgium}
\author{Odemir M. Bruno}
\affiliation{Instituto de Física de São Carlos, Universidade de São Paulo, São Carlos 13566-590, Brazil}
\author{Alexandre S. Martinez}
\affiliation{Faculdade de Filosofia, Ciências e Letras de Ribeirão Preto,
Universidade de São Paulo, Ribeirão Preto 14040-900,
Brazil}
\author{Christophe Deroulers}
\affiliation{Université Paris-Saclay, CNRS/IN2P3, IJCLab, 91405 Orsay, France}
\affiliation{Université Paris-Cité, IJCLab, F-91405 Orsay, France}
\begin{abstract}
Stochastic processes out-of-equilibrium often involve asymmetric
contributions that break detailed balance and lead to non-monotonic
entropy production, limiting thermodynamic interpretations and
inference techniques. Here we use Dyson maps to restore monotonic
entropy growth in those processes, allowing the use of standard tools from statistical
physics, providing a
general and computationally tractable method applicable to a broad
class of Markovian systems.
\end{abstract}
\maketitle

\section{Introduction}
\label{sec:introduction}

Stochastic processes describe the dynamics of systems subjected to 
uncertainties, with wide-ranging applications across physics, biology, 
and finance. In their most elementary formulation, namely Markov 
processes with discrete states (hereafter labelled $i=1, 2, ...$), a 
stochastic system is formalized in terms of a master equation
\begin{equation}
  \label{eq:master}
  \partial_t \lvert P(t) \rangle = - H  \lvert P(t) \rangle,
\end{equation}
which is governed by the time evolution of the probability vector $\lvert P(t) 
\rangle $. The $i$-th component $P_i(t)$ is the probability that the system 
is in state $i$ at time~$t$. The matrix elements $H_{ij}\equiv
H_{ij}(t)$ of the stochastic generator $H$ describe the rates of the  
transitions $j \rightarrow i$, and satisfy the constraint $\sum_i
H_{ij} = 0$ to ensure probability conservation. They are also referred to as 
memoryless transitions since the rates can, at most, depend on the time
instant $t$ acting on $\lvert P(t) \rangle$, i.e., no previous
history is taken into consideration to calculate the probability
distribution at $t+\delta t$. Furthermore, the constraint $ 0
\leqslant P_i(t) \leqslant 1$ also implies that the eigenvalues
$\lambda$ of $H$ satisfy $\textrm{Re}\,{\lambda} \geqslant 0 $.

The operator formulation of Markov processes {
  \cite{stoch-doiJPhysA1976, stoch-grassberger1980, stoch-peliti1985}} has been used in
the past to describe particle diffusion {
\cite{stoch-mattisRevModPhys1998}}, reaction-diffusion processes 
\cite{alcarazAnnPhys1994}, agglutination  and ageing \cite{stoch-henkelJPhysA2004}, 
self-organizing sand-piles \cite{alcarazPhysRevE2008}, motility-induced phase
separation \cite{stoch-nakamuraJStatMech2021}, and epidemics {\cite{stoch-nakamuraSciRep2019,mieghem2009,mieghemPhysRevE2012,epidemics-merbisPhysRevE2023b}} among others.
Often these processes represent ensembles of interacting particles or
individuals, producing collective behaviors and emergent phenomena,
for example the spread of a disease, an idea, or a combination of
them through a population \cite{Castellano2009,Wang2019,VanWesemael2025}. In these cases,  \( H \)
encodes both the characteristics of the contagion and the contact patterns
of the individuals in the population. In the general case, the dimensionality 
of the problem grows exponentially with the number of interacting
particles. Usual solutions include simplifying the system by resorting
to mean-field theories or pair-approximations \cite{epidemics-gleesonPhysRevX2013, ascolaniPhysRevE2013}, with satisfactory 
results when fluctuations are not key or the populations are
large and uniform. {  These exponential scaling issues
  have motivated the use of tensor network and matrix product state
  (MPS) methods to represent high-dimensional operators in compressed
  forms. Originally developed in many-body quantum physics, these
  methods have found their way into stochastic modelling, providing a
  manner to bring  non-trivial microscopic information and assess
  their effects on macroscopic predictions \cite{epidemic-merbisPhysRevE2023,tensor-helmsPhysRevLett2020,stoch-johnsonPhysRevE2010,stoch-henkelEuroPhysJB1999}.
  }


Besides the dimensionality curse, a different issue arises when
dealing with the operator formalism.  
Stochastic generators $H$ are often non-Hermitian. The 
asymmetry arises naturally in systems with
preferred transitions, such as biased random walks, decays, or
irreversible reactions, leading to entropy
production until equilibrium settles. 
However, the entropy production is not always positive for stochastic
processes. As a result, the equilibrium state might encode
configurations with low entropy,  
which undermines techniques based on entropy maximization
\cite{bialekNature2006}. 
Dyson maps address this issue \cite{dyson-moussaPhysicaA2022,dyson-moussaSciPost2023}.
{\color{blue} Dyson maps $\eta$ are reversible transformations that convert non-Hermitian
operators $H$ with real eigenvalue spectra into Hermitian ones,
$\mathcal{H}$, } while keeping the eigenvalues unchanged:
\begin{equation}
  \label{eq:dyson}
  \mathcal{H} = \eta H \eta^{-1}.
\end{equation}
The spectral invariance implies the eigenvalues of
$\mathcal{H}$ belong to $\mathbb{R}$ and
allows for a physical interpretation of the transformed
system. { Similarity transformations have been used in the
past to recover symmetric generators for selected many-body problems
\cite{dyson-henkelAnnPhys1997,stoch-banulsPhysRevLett2019}}.
Here, we demonstrate how to compute Dyson maps in general
settings. Our findings lead to transformed systems in which the
{ entropy production for the Rényi entropy $S_2$} is
always positive, thus imbuing them with a far more familiar physical
interpretation. 
{ Rényi entropies form a family of additive entropies
  characterized by the parameter $q$. The usual 
  Gibb-Shannon entropy is recovered with $q=1$,   $S_1 \equiv S_{\textrm{Shannon}} = -\sum_i
  p_i \ln p_i$,  while $q=
  2$ produces the so-called Rényi entropy or collision entropy $S_2 \equiv
  S_{\textrm{Renyi}} = - \ln\sum_i p_i^2$, with $S_2 \leqslant
  S_1$. 
  In what follows, we use the Rényi entropy as surrogate for entropy
  of the stochastic process as it is calculated directly from the
 scalar product between the probability vector. }
The paper is organized as follows. In Sec.~\ref{sec:dyson-map}, we detail
the properties of Dyson maps,  the interpretation of the
transformed system, and the metric operator $\Omega$, a weight that
connects measurements in transformed and original
coordinates. Sec.~\ref{sec:method} addresses the practical challenge
of finding Dyson maps. 
We emphasize the method
does not rely on stochastic properties and can be extended to general
applications.
Disease spreading models are considered in Sec.~\ref{sec:SIS} to
highlight the role of the Dyson-produced entropy. Further
applications, limitations, and remarks are addressed in Sec.~{\ref{sec:discussion}}.

\section{Dyson maps}
\label{sec:dyson-map}

\begin{figure}[tbh]
  \includegraphics[width=0.9\columnwidth]{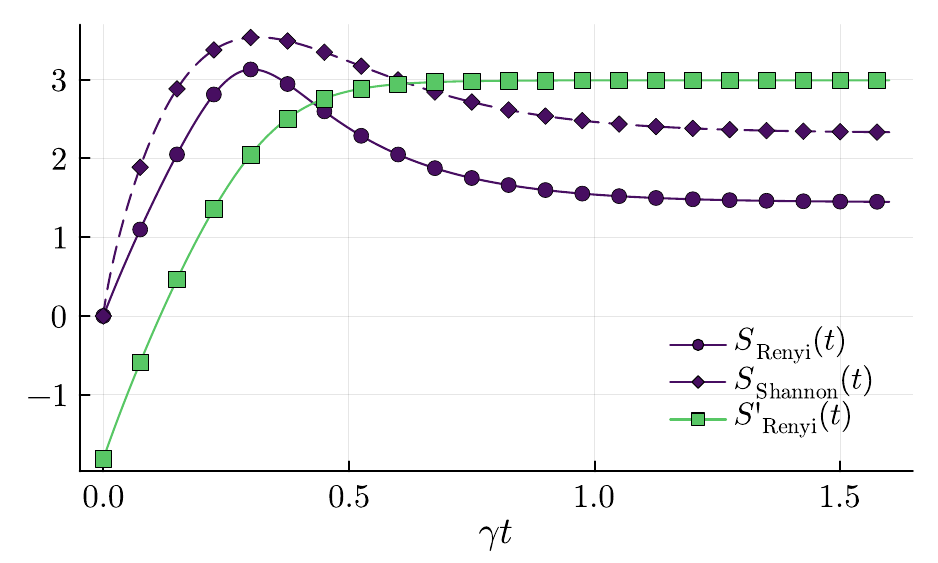}\\
  \includegraphics[width=0.9\columnwidth]{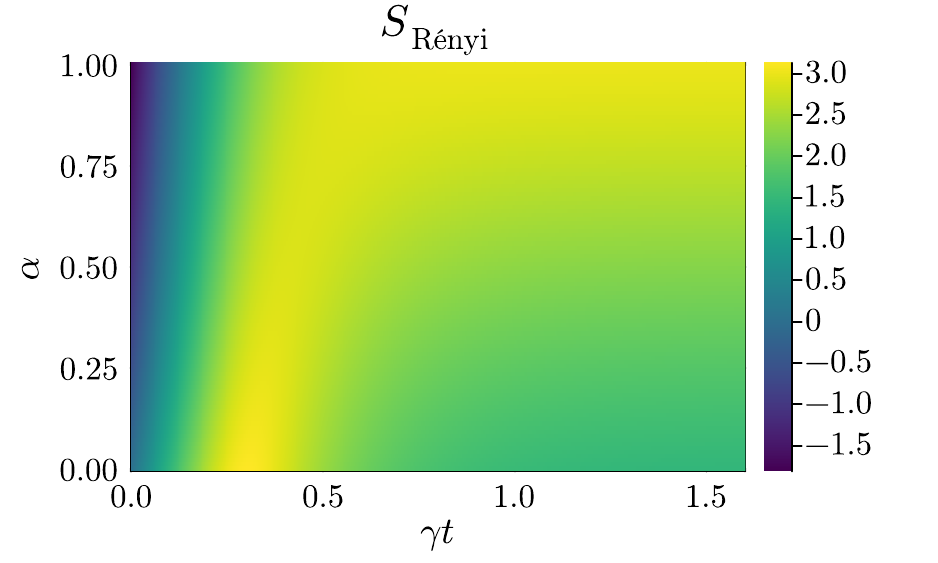}
  \caption{\label{fig:SIS-entropy}
    Time evolution of the SIS-process (\( \rSI/\rIS = 0.1 \)).
    (top) Entropies of the original system $H$ ($S_\mathrm{Renyi}$ and $S_\mathrm{Shannon}$) and for the system $\mathcal{H}$ obtained through Dyson mapping ($S'_\mathrm{Renyi}$).
    (bottom) Renyi entropy along a homotopic map between $\HSIS$ and $\mathcal{\HSIS}$, $((1-\htpar)I + \htpar \eta) H ((1-\htpar)I + \htpar \eta)^{-1}$.
   }
\end{figure}

In what follows, we restrict our analysis to time-independent and
positive semi-definite generators $H$ for the sake of simplicity. In practice,
the subset of operators includes discrete acyclic Markov processes
that relax towards equilibrium in absence of oscillations. Building
from this, we apply $\eta$ on (\ref{eq:master}), which leads to
the following transformed master equation
\begin{equation}
  \label{eq:transformed-master}
  \partial_t \lvert \phi \rangle = -\mathcal{H} \lvert \phi \rangle,
\end{equation}
where the Dyson-transformed vector $\lvert \phi(t) \rangle = \eta
\lvert P(t) \rangle$ lacks a direct interpretation as a
probability vector. To relate the two systems, we investigate
the dynamics of the squared norms of the original probability vector and
of the transformed vector, respectively,
\begin{subequations}
  \begin{align}
    \label{eq:squared-norm-p}
    \partial_t | P(t)|^2 & = -\langle P | H+H^\dag | P \rangle, \\
    \label{eq:squared-norm-phi}
    \partial_t | \phi(t)|^2 & = -2\langle \phi | \mathcal{H} | \phi \rangle.
  \end{align}
\end{subequations}
Recall that the eigenvalues of $\mathcal{H}$ either vanish or are
positive so that $\lvert \phi \rvert^2$ always decreases over time. In
contrast, $(H+H^\dag)$ is in general not semi-positive definite so $\lvert
P \rvert^2$ is not monotonic.

In fact, the local minima of $|P|^2$  occur near regions of maximal uncertainty. This
phenomenon is quite general and suggests a connection with a measure
of uncertainty via the Rényi entropy $S \equiv S_{\textrm{Renyi}} = -\ln
|P(t)|^2 $. However, this association proves somewhat unsatisfactory,
as it is frequently  observed that entropy production is followed by
entropy reduction \cite{nakamuraSciRep2017, melkoPhysRevB2016}.
This fundamental issue arises from the non-monotonic dynamics observed for
$|P(t)|^2$. The cases in which $|P|^2$ decreases monotonically can be
tracked down to Hermitian $H$.

{ In the transformed system, we define the function $S'(t)
  \equiv - \ln |\phi(t)|^2 $ and call it the Rényi entropy for the
  lack of a proper name, recovering the original Rényi
  entropy when $H = H^\dag$. In what follows, $S$ refers to the
  Rényi entropy in the original system, while the prime version $S'$
  refers to the transformed system.}
 The time evolution of $\phi(t)$ implies the
maximization of $S'$  since
\begin{equation}\label{eq:entropy-derivative}
  \frac{1}{2} \frac{d}{d t} S'(t) =   \frac{\langle \phi | \mathcal{H}
    | \phi \rangle}{\langle \phi | \phi \rangle}  \geqslant 0.
\end{equation}
Figure \ref{fig:SIS-entropy} illustrates the phenomenon for an
epidemic SIS process (more details in Sec.~\ref{sec:SIS}), with \(
S'(t) \): the entropy of the original system $S$ reaches a transient
maximum while the transformed one $S'$ increases monotonically until
equilibrium is reached. We claim that Equation
\eqref{eq:entropy-derivative} holds for all Dyson-transformed systems
and, in addition, the  equilibrium states $\lvert\phi_{\textrm{eq}}
\rangle $ are obtained by minimizing $\langle \phi | \mathcal{H} |
\phi \rangle$ or, equivalently, maximizing the entropy $S'$.

In practice, Dyson maps restore microscopic time reversibility while
suppressing stochastic features, in the sense that
$\sum_{i} \mathcal{H}_{ij}\neq 0$ for the transformed system.
For systems away from equilibrium the correct time evolution is
obtained by comparing forward and reverse path weights in the phase
space, $W$ and $W'$ respectively
\cite{nonequ-lebowitzJStatPhys1999}. Violation of time reversibility
means the log-ratio $\log(W/W') \neq 0$, which confers the
time evolution a preferred direction that
might not coincide with the growth of uncertainty. The solution for
this conundrum lies in enforcing that both systems produce the same
value of total entropy over a time interval $\tau$,
specifically,
$\Delta S' = \Delta S + \sum_{\{ W \}} \ln(W/W')$, where the sum is over
all paths $W$ in the phase space which relate the final state $P(\tau)$
to the initial state $P(0)$. We can rewrite
this expression in terms of the metric operator $\Omega =
\eta^\dag\eta$, whose primary role entails re-weighting contributions
$\lvert \phi \rvert^2 = \langle P \rvert \Omega \lvert P \rangle$:
\begin{equation}
  \sum_{\{ W \}}\ln\frac{W}{W'} = - \ln \left[ \frac{\langle P(\tau) | \Omega
      | P(\tau) \rangle}{\langle P(\tau)| P(\tau)\rangle } \, \frac{\langle
    P(0)  | P(0) \rangle}{\langle P(0)| \Omega |P(0)\rangle } \right].
\end{equation}
Thus the total contributions for forward paths can be summarized by
the metric via $  \ln [\langle P(\tau) | \Omega | P(\tau) \rangle \langle
P(0) | P(0) \rangle ]$.


The metric operator has additional statistical applications.
Statistical averages are obtained as follow: $\langle O
\rangle = \sum_{k\ell} O_{k\ell} P_{\ell}(t) = \sum_{k\ell} P_\ell(t) \langle
C_k \rvert O  \lvert C_\ell \rangle$ where $\lvert C_\ell \rangle$ is the vector with all vanishing components except the $\ell$-th which equals 1.
Alternatively, one can explore
$\sum_k\langle P (t) | C_k\rangle = 1$ to rewrite the estimates in a
more familiar form involving the inner product $\langle O \rangle  =
\langle P | \Xi O |P \rangle $, where  $\Xi = \sum_{ij}\lvert
C_i\rangle \langle C_j   \rvert $ samples through every transition and
removes the extra contribution from the left probability vector.  But
since we are dealing with maps, it
is natural to ask how the measurements change since  Dyson maps
are chosen to symmetrize $H$ only. For the general case, given the
operators $A$ and  ${A}' = \eta  A\eta^{-1}$, the following relations
holds: $\langle \phi \vert A' \vert \phi \rangle = \langle P \vert
\Omega  {A} \vert P\rangle$, and $\langle P | A | P \rangle = \langle
\phi \rvert \tilde{\Omega} A' \lvert \phi \rangle$ with the reverse
metric  $\tilde{\Omega} = (\eta\eta^\dag)^{-1}$. Setting $A = \Xi O$
we recover statistical averages:
\begin{equation}\label{eq:statistics}
  \langle O \rangle = \sum_{k \ell} O_{\ell k} P_k(t)  =  \langle \phi
  \vert \tilde{\Omega} {\Xi}' O' \vert \phi \rangle.
\end{equation}
This expression highlights the additional costs to compute statistical
averages in the Hermitian process. The transformation $\eta$ preserves
classical estimates as long as contributions spanning from the metric
or reverse metric are accounted for. This result implies the
advantages to compute the time evolution in an ever increasing entropy
are counter-balanced by additional complexity when computing
statistics encoded by $\tilde{\Omega}$. 

\section{Finding Dyson maps and Hermitian operators}
\label{sec:method}

We want to find $\eta$ that maps the semi-positive
operator $H$ to some corresponding Hermitian operator
$\mathcal{H}$. There is no unique solution to this problem: if $\eta$ is a Dyson map 
transforming $H$ into $\mathcal{H}$, then $U \eta$, where $U$ is any unitary 
matrix, will be another one. Similarly, if $\mathcal{H}$ is a 
transformed operator of $H$, so is $U \mathcal{H} U^{-1}$ for any unitary
matrix~$U$.

We adress this issue in a two steps process. First, solutions $\eta$
related by unitary transformations sharing the same metric operator
$\Omega$ are grouped into so-called orbits. They have the same Rényi
entropy dynamics. Within a given orbit, we seek a Hermitian solution
$\eta =\eta^\dag$ (there is always one); it spans all the remaining
solutions $U \eta$. The second step defines the generator $\Lambda =
\ln \eta = \Lambda^{\dag}$, with matrix representation $d\times d$.
Together with  the traceless Hermitian and skew-Hermitian
contributions, respectively, $\bar{H} \equiv (1/2)(H + H^{\dag}) -
(1/d)\textrm{Tr} H$ and $\Delta H \equiv (1/2) (H - H^{\dag})$, we
seek for $\Lambda$ that satisfies 
\begin{equation}
  \label{eq:comm-1}
  [Q_i, Q_j] = \sum_{k=1}^3 G_{i j k} Q_k,
\end{equation}
where $Q = (\bar{H},\Delta H , \Lambda)$. In addition, it is expected
that $\Lambda \rightarrow 0$  for vanishing $\Delta H$, implying that
$[\bar{H},\Delta H] = G_{122}\Delta H +  G_{123}\Lambda$. Inspired by
early methods  \cite{scholtz1992nonhermitian},  $G_{122}$  is set to
zero { in the so-called orthogonal gauge, where
(\ref{eq:comm-1}) is reinterpreted as a cross-product in the
algebra. The Dyson map $\eta$ thus rotates the system along the axis
$\Lambda$ so that the projection of $\Delta H$ over the time generator
vanishes.

The equation that fixes the remaining coefficients $G_{ijk}$ are
obtained from the identity $\mathcal{H} = \mathcal{H}^{\dag} $ and
expanding it in terms of $\Lambda$:
\begin{equation}
  \label{eq:constraint-1}
  0 = \Delta H + [\Lambda,\bar{H}]+\frac{1}{2!}[\Lambda,
  [\Lambda,\Delta H]] +\frac{1}{3!}[\Lambda,[\Lambda,[\Lambda,\bar{H}]]]+\cdots
\end{equation}
The expression above can be further simplified with use of a matrix
notation for commutators $ [\Lambda, v_1 \bar{H} + v_2 \Delta H +
v_3 \Lambda] \sim \langle \bar{H} , \Delta H , \Lambda\rvert R \lvert v_1,v_2,v_3 \rangle$, with
matrix elements $R_{ij} = G_{3ji}$. The notation compresses commutators
as if they were matrix products and then projects the result onto the
correct operators. It follows that the $n$-th order
commutator $[\Lambda,[\Lambda,[\cdots[\Lambda, v_1 \bar{H} + v_2 \Delta H +
v_3 \Lambda ]\cdots]]] = \langle \bar{H} , \Delta H , \Lambda\rvert
R^n \lvert v_1,v_2,v_3 \rangle$. Therefore, (\ref{eq:constraint-1})
can be re-written as $\langle \bar{H} , \Delta H , \Lambda\rvert \cosh
R \lvert 0,1,0 \rangle + \langle \bar{H} , \Delta H , \Lambda\rvert \sinh
R \lvert 1,0,0 \rangle = 0$. In practice, several coefficients are
obtained by inspection because (\ref{eq:comm-1}) produce $\Lambda$-free equations
under the orthogonal gauge $ [\bar{H},\Delta H] = G_{123} \Lambda$, namely, 
\begin{subequations}
  \begin{align}
    \label{eq:comm-22}
  [[\bar{H},\Delta H], \phantom{\Delta} \bar{H}] &= G_{123}G_{311} \bar{H} +
                                  G_{123}G_{312} \Delta{H} +
                                                   G_{313} [\bar{H},\Delta H],  \\
    \label{eq:comm-23}
    [[\bar{H},\Delta H], \Delta{H}] &= G_{123}G_{321} \bar{H} +
                                  G_{123}G_{322} \Delta{H} +  G_{323} [\bar{H},\Delta H].
\end{align}
\end{subequations}
These equations allow one to recast (\ref{eq:constraint-1}) in simpler
terms to obtain the solution $G_{123}$ and, thus, $\Lambda$.  

The calculation of $\mathcal{H}$ follows from $\mathcal{H} =
(1/2)(\mathcal{H} + \mathcal{H}^\dag)$, 
\begin{align}
  \mathcal{H} & = \frac{1}{d}\textrm{Tr}{H} + \bar{H} + [\Lambda, \Delta
                H] +\frac{1}{2!}[\Lambda,[\Lambda,\bar{H}]] + \frac{1}{3!}
                  [\Lambda,[\Lambda,[\Lambda,\Delta{H}]]] + \cdots \nonumber\\
    \mathcal{H} & = \frac{1}{d}\textrm{Tr}{H} + A_1 \bar{H} + A_2 \Delta H + A_3
                  [\bar{H},\Delta H],
\end{align}
with 
$A_\ell = \delta_{\ell,1} + \sum_{m=1}^{\infty} [ F^{2m}_{1,\ell} +
F^{2m-1}_{2,\ell} ]$ and $F^m_{i,j } = (1/m!)\sum_{\vec{k}=1}^3
\delta_{k_1,i} \delta_{k_{m+1},j} \prod_{r=1}^m G_{3k_r
  k_{r+1}}$. Furthermore, the coefficients satisfy $\textrm{Re}(A_{2})
= \textrm{Im}(A_{1,3}) = 0$.  We detail the process for the decay
problem in the Appendix~\ref{app:decay}.}
{
Before proceeding, we emphasize that the construction above
assumes a closed operator structure as in equation (\ref{eq:comm-1}) for a
Hermitian \(\Lambda\). This requirement is stronger than the minimal
conditions needed for a Dyson map to exist in  general. In particular,
the existence of a set of coefficients $G_{ijk}$ satisfying
(\ref{eq:constraint-1}) ensures the closure relation (\ref{eq:comm-1})
holds  within a finite operator basis. Under this closure, $\Lambda$ and
any similarity transform $U \Lambda U^{-1}$ generate valid Dyson
maps whose action remains within the original operator space. In the
absence of such algebraic closure, similarity transformations can
still be defined at the spectral level, requiring an uncontrolled
number of operators, complicating both practical computations and
interpretation of $\Lambda$.}
  

Although the analytical formulation presented above relies on very few
ingredients and does not specify a specific group or algebra, the
analytical approach remains quite challenging for non-trivial
interacting problems. { Without significant coefficient
reduction by inspection, (\ref{eq:constraint-1}) can produce
infinite solutions, or none at all, without a clear mechanism other
than minimization procedures. From a numerical point of view, the
interpretation of $\Lambda$ as a rotation axis in the algebra provides
a reliable method to reduce $\Delta H$ up to some tolerance. The core
idea steams from choosing iterative rotations around the axis
$\Lambda^{(k)} \propto u
[H^{(k)},\Delta H^{(k)}]$  and then update the main generator $H^{(k+1)}
= \textrm{e}^{\Lambda^{(k)}} H^{(k)} \textrm{e}^{\Lambda^{(k)}}$ at the
$k$-th step. The proportionality coefficient $u^{(k)}$ for $\Lambda^{(k)}$ can be
chosen at random and accepted if $\textrm{Tr}[(\Delta H^{k+1})^\dag\Delta
H^{k+1}] < \textrm{Tr}[(\Delta H^{k})^\dag\Delta H^{k}] $. For large
dimension problems, the exponential operators is replaced by
Trotter-Suzuki formulas preferring small updates $u^{(k)}$ per step.
See Appendix~\ref{app:numerics} for the detailed algorithm, with
publicly available code.
} In what follows,
we apply the numerical method in a toy model to better grasp the
effects of the changes in the transformed system.

\section{Graph-based SIS dynamics}\label{sec:SIS}

\begin{figure}[H]
  \includegraphics[width=0.475\columnwidth]{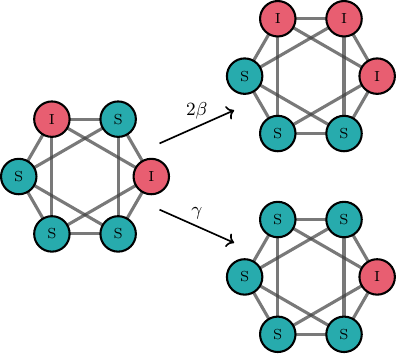}
  \hfill
  \includegraphics[width=0.45\columnwidth]{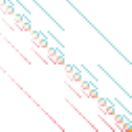}
    \caption{%
        SIS dynamics on a static contact graph.
        The colors indicate the state of the vertices.
        Two possible transitions are given, an infection (top), and a recovery (bottom).
        Right: structure of the stochastic generator $\HSIS$ of the process, with the constant recovery transitions (blue) above the diagonal, the infection transitions (red) below it, indicating the asymmetry of the stochastic process.
    }\label{fig:SIS-model}
\end{figure}

Here, we apply the Dyson map to the system representing a SIS dynamics on a static contact graph.
In this system, there are $\nbverts$ vertices that can be either susceptible or infectious.
An infectious vertex recovers with rate $\rIS$ and infects its susceptible neighbors with rate $\rSI$.
In this case, each state in Eq.~\eqref{eq:master} corresponds to
a specific combination of infectious and susceptible vertices.
We do not consider the infection-free state, as to have at most one
stable state, hence there are $\nbstates = 2^\nbverts - 1$ possible states. 
More concretely the structure of $H$ is given by 
\begin{equation}\label{eq:SIS-H}
    \HSIS = - \rSI \sum_{ij} A_{ij} \ladderup_i\occup_j + \beta \sum_{ij} A_{ij} (1 - \occup_i) \occup_j -\gamma \sum_i \ladderdown_i + \gamma \sum_i \occup_i,
\end{equation}
with $A$ the adjacency matrix of the contact graph, $\occup_i$  the
$i$'th occupation operator and $\ladderup$ ($\ladderdown$) the
operator that invokes an infection (recovery) transition. The decay
parameter can also be recast as an operator $\gamma \rightarrow \gamma
\Theta(\sum_i{n_i} - 1)$ to remove unwanted transitions to the disease-free state,
where $\Theta(x)$ is the Heaviside step function.
{ We examine this system for a ring graph where each vertex is connected to its
four nearest neighbours. Unless otherwise noted, the number of vertices is $\nbverts = 6$.}
Figure \ref{fig:SIS-model} depicts the transitions and the structure of $\HSIS\in\mathbb{R}^{63\times63}$.
The upper triangular part the represents the recoveries (all with rate
$\rIS$), while the lower triangular part contains the infections,
where the rate is a multiple of $\rSI$, depending on the number of
infectious neighbors. In what follows, we study the Hermitian $\hSIS$
using the algorithm described in Appendix \ref{app:numerics}.
Because for some combinations of \( \nbverts \) and \( \rSI/\rIS \),
eigenvalues of \( \HSIS \) may have small imaginary parts, we minimally
perturb the system to one with real spectrum.

\begin{figure}[H]
    \centering
    \includegraphics[width=0.9\columnwidth]{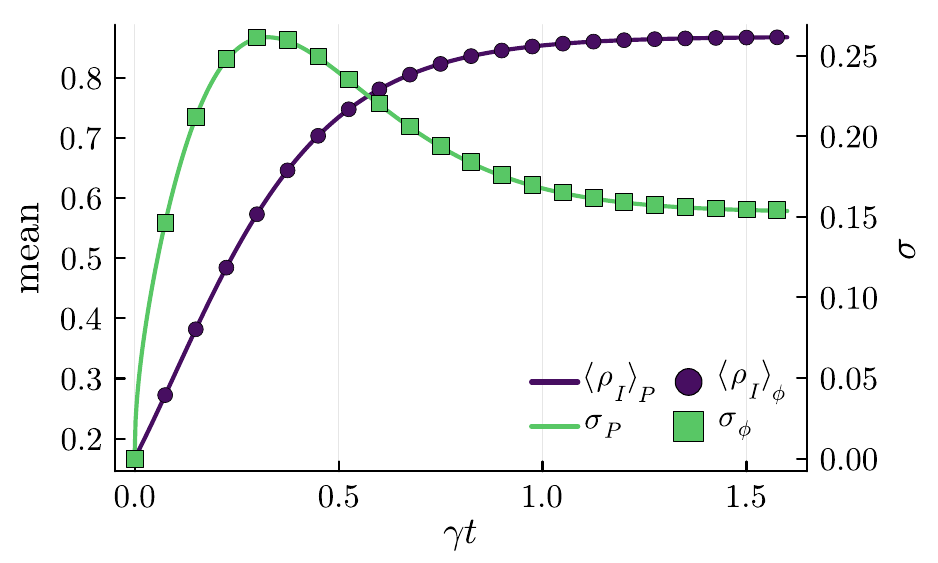}
    \caption{Time evolution of the mean $\langle \rho_I \rangle$ and standard
      deviation $\sigma$ of the proportion of infectious
      vertices in the SIS system (for \( \rSI/\rIS = 0.1 \)), obtained by
      integration of $\HSIS$ (lines) and $\hSIS$ (markers).}
    \label{fig:SIS-statistics}
\end{figure}

We consider the time evolution of several statistics for
$\rSI=10^{-3},\rIS=10^{-2}$, starting from one infectious vertex chosen at random.
Figure \ref{fig:SIS-statistics} shows the dynamics for the mean and
the standard deviation of the proportion of infectious vertices  for
$\HSIS$ (lines) and $\hSIS$ (markers).
We use the formula in Eq.~\eqref{eq:statistics} to ensure the
correspondence between the statistics in both original and transformed
system, with excellent  agreement for the numerical results.
Initially, as the infection starts spreading, the standard deviation
increases, but when the mean proportion is approaching the steady
state, it decreases again resulting in a non-monotonous curve.
The evolution of entropy in Figure \ref{fig:SIS-entropy} confirms this behavior.
During the spreading phase, the Shannon and Rényi entropies reach a maximum for
the original system, but settle in the equilibrium state.
In contrast, for the transformed $\hSIS$, the Rényi entropy increases monotonically
through Eq.~\eqref{eq:entropy-derivative}.

Figure \ref{fig:SIS-entropy} (bottom) exhibits the changes of the
Rényi entropy as it continuously transforms from  $\HSIS$ to
$\hSIS$ through the means of the homotopic map with $\alpha \in [0,1]$:
\begin{equation}
  \eta(\alpha) = (1-\alpha)I + \alpha \eta_{\textrm{true}},
\end{equation}
{ where $\eta_{\textrm{true}}$ is a correct Dyson map for $H_{\textrm{SIS}}$.}
The entropy profile evolves continuously, with the maximum shifting
and broadening as $\htpar$ increases. In practice, the entropy peak in
the original system encodes the information necessary to construct the
equilibrium state in the transformed system. This result supports the
educated guess that dynamical features associated with transient regimes
play a critical role in models of disease spreading.

\begin{figure}
  \includegraphics[width=0.9\columnwidth]{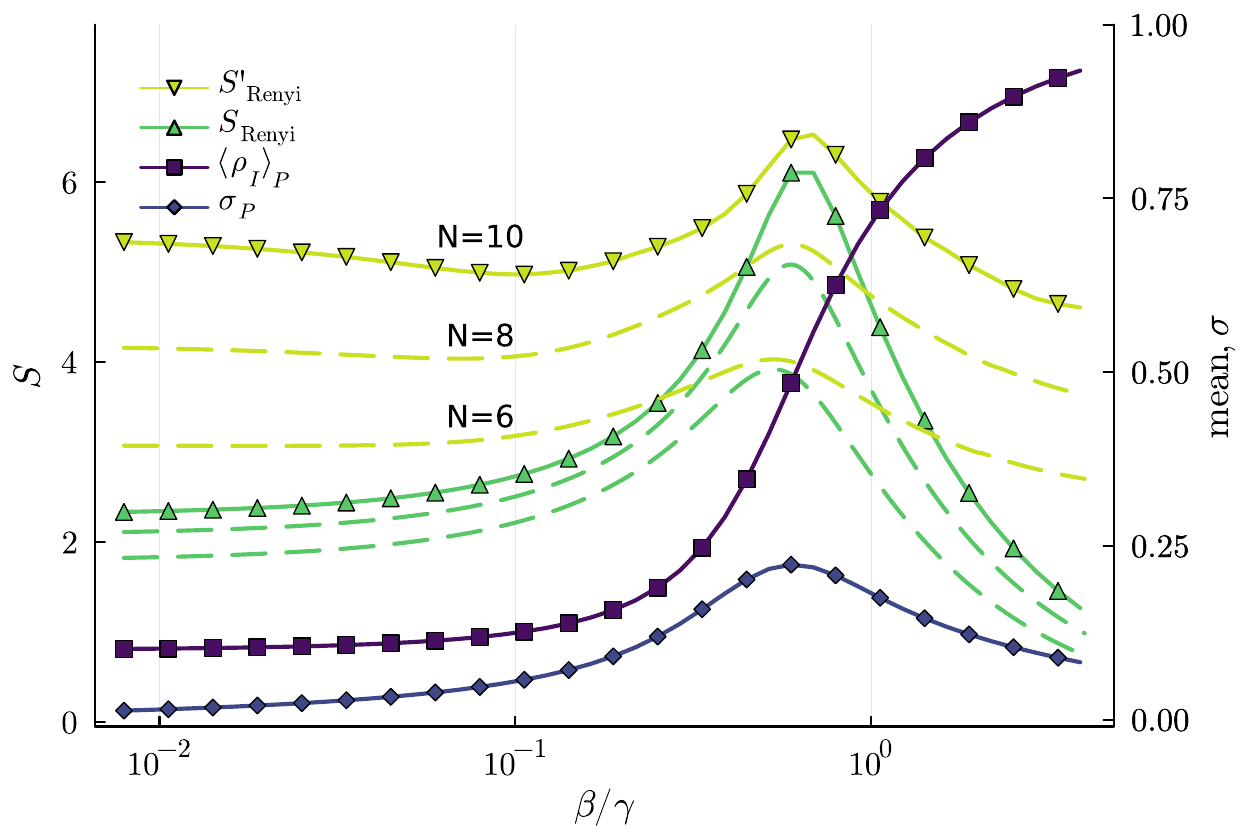}
  \caption{\label{fig:SIS-transition}
  Phase diagram of the SIS system. In equilibrium, average and standard 
  deviation of the proportion of infectious vertices in the larger 
  system, as well as Rényi entropy of original and transformed systems 
  of different sizes \( \nbverts \) as a function of the ratio $\rSI/\rIS$.}
\end{figure}

\begin{figure}
  \includegraphics[width=0.66\columnwidth]{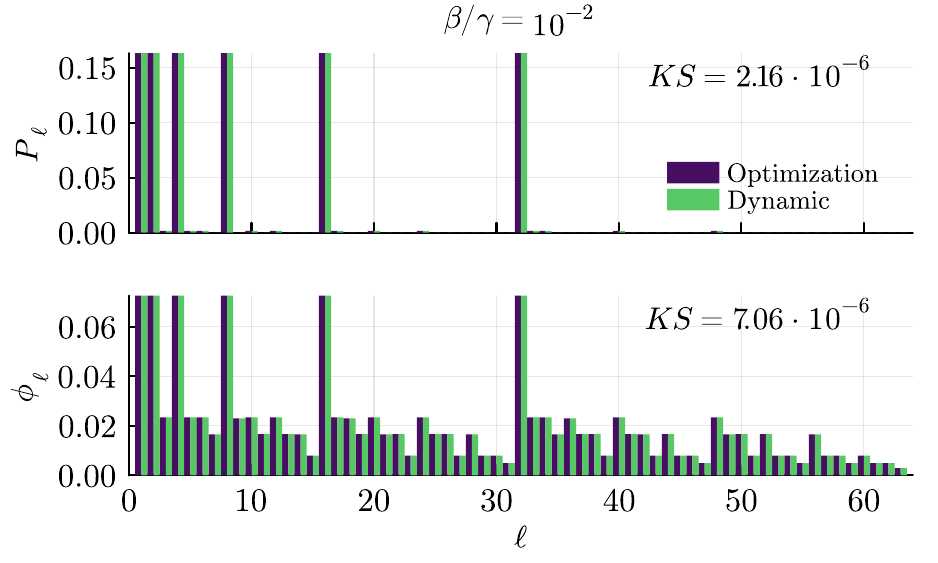}
  \includegraphics[width=0.66\columnwidth]{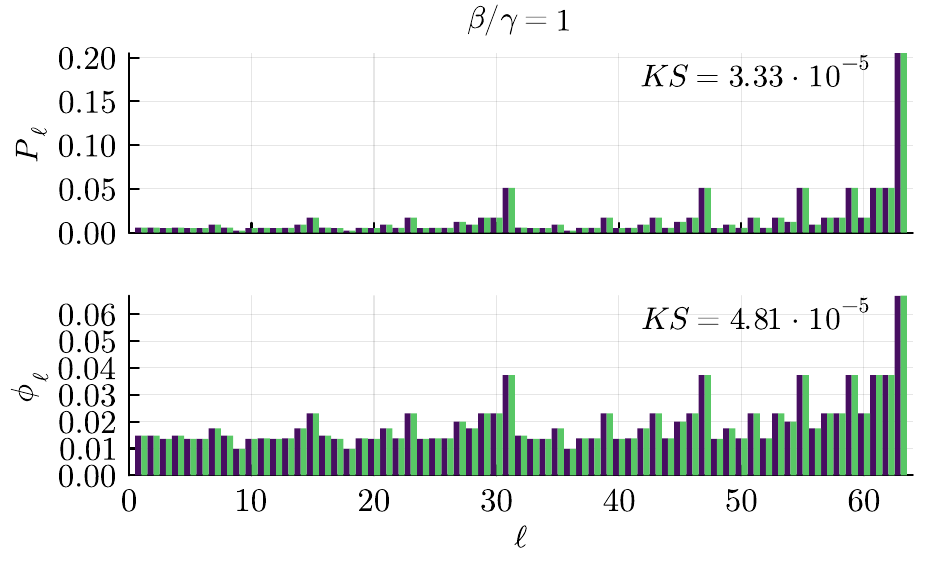}
  \includegraphics[width=0.66\columnwidth]{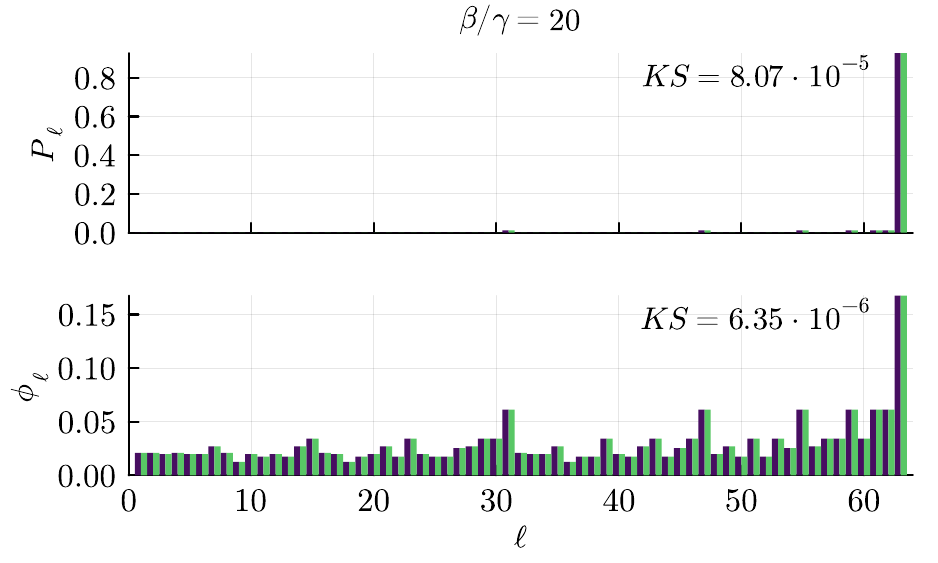}
  \caption{\label{fig:SIS-steady-state}
    Values of $P_\ell$ and $\phi_\ell$ in the steady state for
    $\rSI/\rIS=10^{-2}$ (top), $\rSI/\rIS=1$ (middle) and $\rSI/\rIS=20$ (bottom), obtained by solving Eq.~\eqref{eq:master} (green) and Eq.~\eqref{eq:entropy-derivative} (purple), along with the Kolmogorov–Smirnov ($KS$) distance between them.
  }
\end{figure}

Finally, we study the phase transition in the SIS process.
Figure~\ref{fig:SIS-transition} shows the equilibrium proportion
$\langle \rho_I \rangle$ of infectious individuals, which rises
from zero to one as the ratio $\rSI/\rIS$ increases.
For $\rSI/\rIS \ll 1$, the system reaches a steady state where
configurations with a single infectious vertex are equally likely. The
remaining configurations acquire non-trivial likelihood for moderate
ratios $\rSI/\rIS$, resulting in endemic outbreaks and a sharp
increase of entropy.
{ In both regimes, the two total amounts of produced entropy
  $\Delta S'$ and $\Delta S$ are different, indicating that the original
  steady state violates microscopic reversibility.
In contrast, the values of the ratio $\rSI/\rIS$ for which the Rényi 
entropy is maximal are the same in both formulations ( \( 
S_\mathrm{Renyi} \) and \( S'_\mathrm{Renyi} \)), and this is even true 
at each finite system size $N$. Now, this maximum of entropy is a 
signature of the phase transition in the SIS model between phases where 
the proportion of infectious vertices in equilibrium is below resp. over 
one half, as can be seen, with some finite-size effects, by comparing 
curves for the entropy and for the proportion of infectious vertices. 
Therefore, the transformed system, where time reversibility has been 
restored and where the Rényi entropy \( S'_\mathrm{Renyi} \) 
monotonically increases with time, can be used for the detection and 
location of the phase transition.}

Figure~\ref{fig:SIS-steady-state} allows us to examine these findings
more closely. It shows the steady-state values of  $P_\ell$ and
$\phi_\ell$  before, during, and after the phase transition.
These were obtained both by numerically integrating the master
equation Eq.~\eqref{eq:master}, and by solving the optimization
problem Eq.~\eqref{eq:entropy-derivative}. We employed the
Kolmogorov-Smirnov distance to quantify the agreement between their
respective distributions. In all three regimes, the two solutions
match to within $10^{-3}$, with the largest discrepancy occurring near the phase
transition. An inspection of $|P\rangle$ for increasing $\beta/\gamma$
reveals how the steady state evolves from configurations with a
single infectious vertex to a distribution with a wider support,
eventually converging to the fully infectious state for $\beta/\gamma
\gg 1$. Remarkably, in the transformed state, $\lvert \phi \rangle$, not
only has increasing entropy with time, as shown in
Fig.~\ref{fig:SIS-entropy} and
Fig.~\ref{fig:SIS-transition}, but also conserves some of the
structure of $|P\rangle$. Specifically, the indices $\ell$ with the
highest values of $P_\ell$ and $\phi_\ell$ coincide across all three
regimes. This suggests that diagonal contributions $\eta_{\ell\ell} $
govern the transformation, whereas  off-diagonal components actively
mix components to increase the Rényi entropy.

\section{Conclusion}
\label{sec:discussion}

Stochastic systems are ubiquitous in data analysis and modelling in 
natural and social sciences. They encode the
dynamics of quantities of interest, together with the effects created by
fluctuations. In general, the complete treatment of general stochastic
process can be simplified under certain assumptions. For instance,
compartmental epidemiological models with $m$ states are described by
$m$ equations, instead of $m^{2N}$, under the key assumption of
statistical independence. For finite or non-trivial noise,  one must
consider the full problem, leading to entropy production in a setting
away from equilibrium. In this 
context, the direct use of entropy measures often leads to scenarios
where the entropy production becomes negative, contrasting with our
naive expectation of an ever growing entropy, and thus a time
arrow. 
Instead, the lack of time reversibility significantly
amplifies uncertainty production, captured by log-ratios for the
forward and backward paths in the phase space. 

Dyson maps restore temporal reversibility, transforming a
non-Hermitian stochastic process into a Hermitian model with a
physical interpretation. { In other words, the transformed 
system acquires new transitions as if subjected to new interactions. For 
instance, the decay problem becomes a spin system under magnetic 
fields, from which one can calculate, with support of the metric, any 
stochastic observable from the original system (see 
Appendix~\ref{app:decay}).} Our approach builds from rotations $\Lambda$
in the algebra with goal to suppress non-Hermitian contributions. 
Our findings also show that such systems
have a strictly positive entropy production, and the equilibrium state
can be calculated by traditional optimization procedures. This feature
opens up venues to calculate statistics for very slow processes or to
combat critical slowing down in numerical simulations. 



Following our approach, a formal connection emerges between
stochastic processes and quantum  
dynamics through a Wick rotation of time if the eigenvalues
$\{\lambda\} \in \mathbb{R}$ and $\lambda \geqslant 0$.  
Namely, Eq.~(\ref{eq:master}) takes the same mathematical form as a
Schrödinger equation via $t \rightarrow i\tau$. The connection becomes
clear if one recalls that the time evolution of the probability
densities occurs simultaneously across all configurations. Of course,
there are subtle differences that arise due to the manner that
probabilities are calculated in both formulations. The coefficients
$P_\ell(t) $ are probabilities in stochastic processes, in contrast to the
squared norm in quantum  systems. This difference excludes all the
quantum effects related to superposition, and also
ties the probability interpretation in the stochastic process to a
single basis. The analogy suggests quantum simulators 
can be used to tackle hard day-to-day classical stochastic systems
where noise and fluctuations take a prominent role in hard-to-solve
problems \cite{epidemics-wangPhysRevApplied2023}.

Finally, our approach emphasizes the role of pseudo-Hermitian
operators, ie, strictly real spectra. While these operators form a
substantial family in stochastic processes, imaginary eigenvalues are
expected in problems with some degree of oscillations. They become
even more relevant in finite systems and can lead to characteristic
times, thus very relevant for data analysis. Our approach cannot
address complex spectra due to the imposition of Hermiticity. Ideally,
this condition could be relaxed and, instead, one would seek for
transformations with degenerate contributions from states with
conjugated eigenvalues. This is already the case for regular
stochastic processes but it is not clear at this time how to restore
time reversibility for these systems.



\begin{acknowledgments}
  This work was jointly supported by FAPESP (2023/07241-5,
  2021/08325-2), CNPq (305610/2022-8) and FWO (G0G0122N).
  GN thanks the hospitality of the University of Ghent and of IJCLab,
  Univ. Paris-Saclay, Paris Cité and CNRS/IN2P3, where part of this
  work was carried out.
  ASM acknowledges Brazil’s  National Council for Scientific and
  Technological Development CNPq  (grant no. 0304972/2022-3) and the
  financial support by National Institute of Science and Technology in
  Innovative Research in Health Sciences  from Nanotechnology to
  Artificial Intelligence (INCT PICS) CNPq, grant no. 408417/2024-2,
  Coordination of Superior Level Staff Improvement (Capes), grant
  no. 88887.197686/2025-00, and S\~ao Paulo Research Foundation
  (FAPESP), grant no. 2025/26818-7.
\end{acknowledgments}

\appendix
\section{Decay process}
\label{app:decay}

{
Consider the 2-level decay process $\lvert 1 \rangle
\xrightarrow{\alpha} \vert 0 \rangle$ with rate $\alpha > 0$. The
stochastic matrix reads
\begin{equation}
  H = \left(
  \begin{array}{cc}
    +\alpha & 0 \\
    -\alpha & 0
  \end{array}
  \right).
\end{equation}
Here $\bar{H} = (\alpha/2)(\sigma_3 - \sigma_1)$ and $\Delta
H = i(\alpha/2)\sigma_2$, where $\sigma_{1,2,3}$ are the usual Pauli
matrices.  The commutators are  
\begin{subequations}
  \begin{align}
    \label{eq:decay1}
    [\bar{H},\Delta H] & =
                        (1/2) \alpha^2 (\sigma_3+\sigma_1)
                          \\    
    [[\bar{H},\Delta H],\phantom{\Delta}\bar{H}]   &=  -2\alpha^2 \Delta H, \\
    [[\bar{H},\Delta H],\Delta{H}] &= - \alpha^2  \bar{H}. 
  \end{align}
\end{subequations}
From (\ref{eq:comm-1}), one identifies $G_{311} = G_{322} = G_{313} =
G_{323} = 0$, with $G_{321} =  (1/2) G_{312} = - \alpha^2 /
G_{123}$. Putting these coefficients in (\ref{eq:constraint-1})
produces an equation that only contains $\Delta H$,
vanishing for $\tanh(\sqrt{2}G_{321}) = - 1/\sqrt{2}$. When compared to
(\ref{eq:decay1}) we finally obtain 
\begin{subequations}
  \begin{align}
    \Lambda & = \frac{1}{2\sqrt{2}}\tanh^{-1}\left(\frac{1}{\sqrt{2}}\right) (\sigma_1 +
    \sigma_3), \\
\mathcal{H} & = \frac{ \textrm{Tr}(H) I + \sqrt{2}\bar{H} }{2}. 
\end{align}
\end{subequations}
For the 2-level decay, $\Lambda$ does not depend on the transition rate $\alpha$. In general, the map
$\eta$ will be a function of the various transition rates involved in 
the stochastic process.
}


\section{Numerical procedure}\label{app:numerics}

\begin{figure}
  \includegraphics[width=0.9\columnwidth]{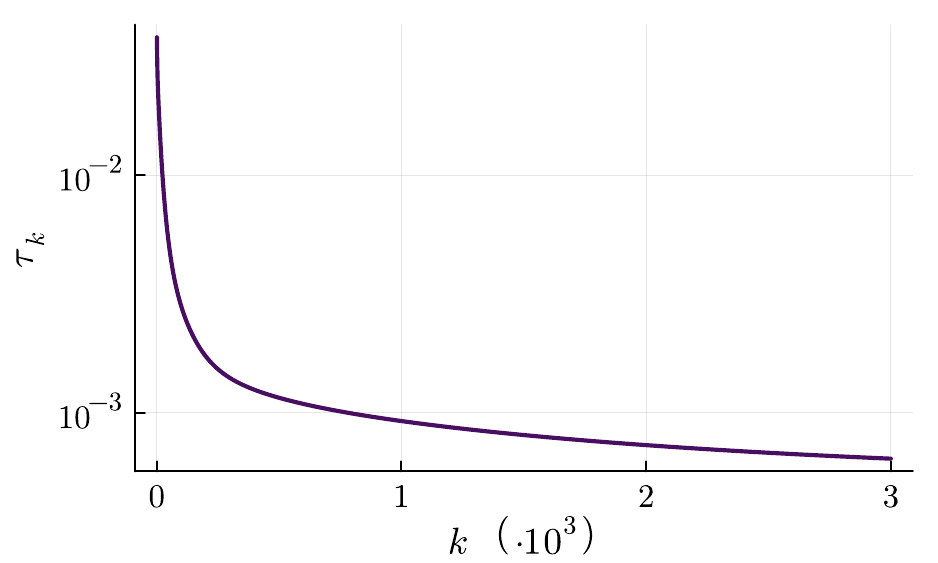}  \caption{\label{fig:convergence}
    Convergence of Algorithm~\ref{alg:dyson} for the SIS model.
  }
\end{figure}

To support and illustrate our analytical results, we use the simple numerical Algorithm~\ref{alg:dyson} to find Dyson maps.
We present it, without any claims regarding convergence behavior, or computational cost.
The idea is to iteratively construct a higher dimensional subspace of the algebra, and find an optimal step size in the added direction.
Figure \ref{fig:convergence} shows the convergence behavior for the SIS-model in Section~\ref{sec:SIS}.

\begin{algorithm}[H]
    \caption{Find a Dyson map for $H_0$ with tolerance $\epsilon$ in maximum $K$ iterations}\label{alg:dyson}
    \begin{algorithmic}
        \Require $H_0$, $K$, $\epsilon$
        \State $k \gets 0$
        \State $\tau_0 \gets \infty$
        \While{$\tau_k > \epsilon$ \textbf{and} $k<K$}
            \State $\Delta H \gets \frac{1}{2}(H_k - H_k^\dagger)$
            \State $A \gets [H_k, \Delta H] = -[H_k, H_k^\dagger]$
            \State $A \gets A / \|A\|_F$
            \State $a \gets \arg\min_x \left\| \exp(xA) H \exp(-xA) - \left(\exp(xA) H \exp(-xA)\right)^\dagger \right\|_F$
            \State $H_{k+1} \gets \exp(aA) H \exp(-aA)$
            \State $k \gets k + 1$
            \State $\tau_k \gets \left\| H_k - H_k^\dagger \right\|_F / n$
        \EndWhile
        \State $H \gets \frac{1}{2}(H_k + H_k^\dagger)$
    \end{algorithmic}
\end{algorithm}

%

\end{document}